\documentclass[12pt,report]{elsart}

\usepackage{hyperref}
\usepackage{epsfig,graphics,colordvi}
\usepackage{amsmath}
\usepackage{amssymb}
\usepackage{mciteplus}

\begin{document}
\begin{frontmatter}
\Large{\title{ Dynamical light vector mesons in low-energy scattering of Goldstone bosons}}
\author[GSI,ITEP]{I.V. Danilkin,}
\author[GSI]{L.I.R. Gil,}
\author[GSI]{and M.F.M. Lutz}
\address[GSI]{GSI Helmholtzzentrum f\"ur Schwerionenforschung GmbH,\\
Planck Str. 1, 64291 Darmstadt, Germany}
\address[ITEP]{Institute for Theoretical and Experimental Physics,\\
117259, B. Cheremushkinskaya 25, Moscow, Russia}
\begin{abstract}

We present a study of Goldstone boson scattering based on the flavor SU(3) chiral Lagrangian formulated
with vector mesons in the tensor field representation. A coupled-channel computation is confronted
with the empirical s- and p-wave phase shifts, where good agreement with the data set is obtained up to about 1.2 GeV. There are two relevant free parameters only, the chiral limit value of the pion decay constant and the coupling constant characterizing the decay of the rho meson into a pair of pions. We apply a recently suggested approach that
implements constraints from micro- causality and coupled-channel unitarity. Generalized potentials are obtained
from the chiral Lagrangian and are expanded in terms of suitably constructed conformal variables. The partial-wave
scattering amplitudes are defined as solutions of non-linear integral equations that are solved by means of an
N/D ansatz.

\end{abstract}
\end{frontmatter}

%\tableofcontents
\section{Introduction}

The study of Goldstone boson interactions is a time honored
challenge in hadron physics
\cite{vanBeveren:1986ea,Weinstein:1990gu,Janssen:1994wn,Roy:1971tc,Ananthanarayan:2000ht,Colangelo:2001df,Buettiker:2003pp,GarciaMartin:2011cn}.
In the last decade there was significant progress in a profound
description and understanding of the low-energy regime based on
the chiral Lagrangian. In the close-to-threshold region chiral
perturbation theory ($\chi$PT) is applicable and leads to
controlled results that are consistent with the empirical
scattering data
\cite{Weinberg:1978kz,Gasser:1983yg,Knecht:1995tr,Bijnens:1995yn,Bijnens:1997vq}.
Extensions of $\chi$PT, that implement coupled-channel unitarity
by means of partial summation techniques, aim at a description of
the scattering data at somewhat larger energies where meson
resonances play an important role
\cite{Oller:1997ti,Bernard:1991zc,Jamin:2000wn,GomezNicola:2001as}.
Most remarkable is the systematic computation at the one-loop
level where the inverse of the partial-wave amplitudes is expanded
in chiral powers \cite{GomezNicola:2001as}. Accurate results for
the s- and p-wave phase shifts up to about $\sqrt{s} \simeq $ 1.2
GeV were obtained by adjusting the $Q^4$ counter terms of the
chiral Lagrangian. This implies that the lowest scalar and vector
mesons with $J^P = 0^+$ and $1^-$ are properly described by such
an approach. It has been argued by detailed large-$N_c$ scaling
studies that the nature of the light scalar and vector mesons is
distinct \cite{Lutz:2008km}. While the scalar mesons may be
generated by the leading order terms of the chiral Lagrangian, the
vector mesons are a consequence of the subleading counter terms.

The purpose of this Letter is to further explore the dynamic role
of light vector mesons in a chiral Lagrangian. The motivation of
considering the light vector mesons as explicit degrees of freedom
is twofold. First, we recall that it was shown previously that the
size of the symmetry preserving $Q^4$ counter terms may be
estimated by a resonance saturation mechanism where the light
vector mesons appear to dominate \cite{Ecker:1988te}. Second, the
light vector mesons play a particular role in the hadrogenesis
conjecture \cite{Lutz:2008km,Lutz:2003fm}. Together with the
Goldstone bosons, they are identified to be the relevant degrees
of freedom that are expected to generate the meson spectrum. For
instance it was shown that the leading chiral interaction of
Goldstone bosons with the light vector mesons generates an
axial-vector meson spectrum that is quite close to the empirical
one \cite{Lutz:2003fm}. From this point of view the axial-vector
mesons are similar in nature as the scalar mesons. We consider a
chiral Lagrangian with explicit vector meson fields as a promising
starting point to compute the meson spectrum.

While it is straightforward to write down a chiral Lagrangian with explicit vector mesons it is an open issue how to
order the various terms according to some counting scheme. In a recent work a counting scheme was suggested based on
large-$N_c$ arguments and the dynamical assumption of hadrogenesis \cite{Lutz:2008km}. So far it has been tested successfully
against two and three body decays systematically at the tree level \cite{Lutz:2008km,Leupold:2008bp}. In this work we scrutinize its implication
in the scattering of Goldstone bosons. At leading order the vector meson exchange processes should be considered
already, rather than at a subleading order, as is implicit in the standard $\chi$PT approach. Based on this assumption
we perform a coupled-channel computation, where we apply a novel unitarization scheme developed by one of the authors
recently \cite{Gasparyan:2010xz,Danilkin:2010xd}.

Though the inverse amplitude method applied with great success in previous works obtains an inverse amplitude that
can be represented in terms of dispersion-integral representation there is a subtle limitation of this approach, if
applied to a coupled-channel situation. In the inverse amplitude method or any algebraic approach the partial-wave
scattering amplitudes have unphysical left-hand branch points. This holds at any finite truncation. The source of these
is easily understood. The locations of the left-hand branch points of a partial-wave scattering amplitude depend
decisively on the channel. There is no universal branch point that characterizes all coupled-channel amplitudes. In
any algebraic method there is a determinant of functions with left-hand branch points at different locations. Via the
determinant a branch point of a given channel is transported into any other channel as long as the transition potential
is non-vanishing. In general this leads to the presence of all possible left-hand branch cuts in all channels. Though
this is unphysical it does not necessarily always lead to numerically significant effects in the physical region. If all
considered branch points are below the smallest considered threshold typically the presence of unphysical branch points
are not really problematic. However, once a left-hand branch point of a heavy channel is located right to the threshold
of a lighter channel the limitations of an algebraic approach start to be visible.

The main objective of the novel scheme \cite{Gasparyan:2010xz,Danilkin:2010xd}  to be applied in this work
is a controlled realization of the causality and unitarity condition in a perturbative application of the chiral
Lagrangian. The starting points are partial-wave dispersion relations. A generalized potential is constructed from
the chiral Lagrangian in the subthreshold region and analytically extrapolated to higher energies. The partial-wave
scattering amplitudes are obtained as solutions of non-linear integral equations.

In our scheme the empirical s- and p-wave phase shifts are recovered  up to about 1.2 GeV. There are two relevant
free parameters only, the chiral limit value of the pion decay constant and the coupling constant characterizing the
decay of the rho meson into a pair of pions.

\section{Chiral Lagrangian with vector meson fields}
\label{Chiral Lagrangian}

We recall the  leading order chiral Lagrangian for the  Goldstone
bosons and vector mesons. The vector mesons are represented in
terms of anti-symmetric fields $V_{\mu \nu} = - V_{\nu \mu}$. It
has been demonstrated first in \cite{Ecker:1988te}, that vector
mesons interpolated in terms of anti-symmetric fields provide a
transparent saturation model for the $Q^4$ counter terms of the
chiral Lagrangian in its Goldstone boson sector. The chiral
Lagrangian is composed out of the building blocks, $U_\mu,
\,V_{\mu \nu}, \,\chi_\pm$,  that  transform identically under
chiral  rotations (see e.g. \cite{Krause:1990}). We do not
consider electromagnetic interactions here. Using covariant
derivatives, $D_\mu$, the transparent transformation properties
are maintained for terms involving derivatives,
\begin{eqnarray}
&&D_\mu \,V_{\alpha \beta}  = \partial_\mu\,V_{\alpha \beta} +\big[\Gamma_\mu ,V_{\alpha \beta}\big]_- \,, \qquad \qquad  \quad \;\;\,
\Gamma_\mu = \frac{1}{2}\,\Big( u^\dagger \,\partial_\mu\,u +u
\,\partial_\mu\,u^\dagger \Big) \,,
\nonumber\\
&& U_\mu = \frac{1}{2}\, u^\dagger \,\Big(\partial_\mu e^{i\,\frac{\Phi}{f}}\Big) \,u^\dagger  \,, \quad
u = \exp \left( \frac{ i \,\Phi }{2\,f} \right)\,, \quad
\chi_\pm  = \frac{1}{2}\,u\,\chi_0 \,u \pm \frac{1}{2}\,u^\dagger\,\chi_0 \,u^\dagger \,,
\nonumber\\
\end{eqnarray}
\begin{eqnarray} && \chi_0 = 2\,B_0 \, \left(
\begin{array}{ccc}
m_u & 0 & 0\\
0 & m_d & 0 \\
0 & 0 & m_s
\end{array}
\right) =\left( \begin{array}{ccc} m_\pi^2 & 0 & 0\\ 0 & m_\pi^2 &
0 \\ 0 & 0 & 2 \,m_K^2 - m_\pi^2 \end{array} \right)\,,
\label{eq:defcovder}
\end{eqnarray}
where the pseudo-scalar meson octet field $\Phi$ is normalized as
in \cite{Lutz:2008km}. There are two types of terms which one has
to consider. The ones that break the chiral SU(3) symmetry
explicitly involve the field combinations $\chi_\pm$ and are
proportional to the quark-mass matrix of QCD. At leading order the
diagonal matrix $\chi_0$ can be expressed in terms of the pion and
kaon masses as indicated in (\ref{eq:defcovder}). Isospin breaking
effects are neglected. We consider further constraints from QCD as
they arise in the limit of a large number of colors $N_c$
\cite{Lutz:2008km}. This implies that interaction terms that
involve a double trace in flavor space can be neglected. The
latter are suppressed by $1/N_c$ as compared to single-flavor
trace interactions. To order $Q^2$ the following terms were
proposed in \cite{Lutz:2008km} to be relevant in our case
\begin{eqnarray}
\label{kinetic-term} {\mathcal L} &=& f^2\, {\rm tr} \Big\{
U^\mu\,U^\dagger_\mu  +\frac{1}{2}\,\chi_+ \Big\}\,
-\frac{1}{4}\,{\rm tr}\, \Big\{(D^\mu\,V_{\mu \alpha})\,(D_\nu
\,V^{\nu \alpha})\Big\}
\\
\nonumber &+& \frac{1}{8}\,m_{1^-}^2\,{\rm tr} \,\Big\{ V^{\mu
\nu}\,V_{\mu \nu}\Big\} +\frac{1}{8}\,b_D\,{\rm tr} \,\Big\{V^{\mu
\nu}\,V_{\mu \nu} \, \chi_+ \Big\}+i\,\frac{m_V\, h_{P}}{2}\,{\rm
tr}\,\Big\{U_\mu\,V^{\mu\nu}\,U_\nu\Big\}
\end{eqnarray}
where  $f \simeq f_\pi$ may be identified with the pion-decay
constant, $f_\pi =92.4$ MeV, at leading order. A precise
determination of $f$ requires a chiral SU(3) extrapolation of some
data set. In \cite{Lutz:2001yb} the value $f \simeq 90$ MeV
obtained from a detailed study of pion- and kaon-nucleon
scattering data. This value was used consistently in various
applications of chiral Lagrangians to meson and baryon resonance
physics \cite{Lutz:2001dr,Lutz:2004dz,Lutz:2005ip}. We
discriminate the mass of the light vector mesons $m_{1^-}$  and
the scale parameter $m_V$. From \cite{Lutz:2008km,Leupold:2008bp}
we recall the value for the parameter, $h_P$, as determined by two and
three-body decay properties of the light vector mesons
\begin{equation}
h_P = 0.29\pm 0.03\,  \quad \textrm{for} \quad m_V=776\, {\rm MeV}
\label{collection-of-parameters}
\end{equation}
at tree-level computations. A first coupled-channel computation
based on (\ref{kinetic-term}), was presented in \cite{Lutz:2003fm}
where the scattering of Goldstone bosons off the vector mesons was
considered. The axial-vector meson spectrum was generated
dynamically.

\section{Coupled-channel dynamics}
\label{Coupled channel}

Given the Lagrangian (\ref{kinetic-term}) we compute the two-body coupled-channel scattering amplitudes. At tree-level
the scattering amplitude takes the generic form
\begin{eqnarray}
&& T(s,t,u)=\frac{C_{s}} {12f^2}\,s+\frac{C_{t}} {12f^2}\,t+\frac{C_{u}}{12f^2}\,u
+\frac{C_{\pi}}{12f^2}\,m_{\pi}^2+\frac{C_{K}}{12f^2}\,m_{K}^2
\label{Tstu}\\
&& \!\!\!\!\!\!\!\!\!+\sum_{x=0}^{8}\frac{(m_{V}h_{P})^2}{128f^4}\frac{C^{(x)}_{s-ch}}{s-m_x^2}\left[(s + m_{2}^2
- m_{1}^2) (s +  \bar m_{2}^2 -  \bar m_{1}^2) + 2\, (\,t - m_{2}^2 -  \bar m_{2}^2)\, s\,\right]
\nonumber\\
&& \!\!\!\!\!\!\!\!\!+\sum_{x=0}^{8}\frac{(m_{V}h_{P})^2}{128f^4}\frac{C^{(x)}_{t-ch}}{t-m_x^2}\left[ (\,t + m_{2}^2
-  \bar m_{2}^2) (\,t + m_{1}^2 -  \bar m_{1}^2) + 2\, (s - m_{2}^2 - m_{1}^2)\; t\,\right]
\nonumber\\
&& \!\!\!\!\!\!\!\!\!+\sum_{x=0}^{8}\frac{(m_{V}h_{P})^2}{128f^4}\frac{C^{(x)}_{u-ch}}{u-m_x^2}
\left[(u + m_{2}^2 -  \bar m_{1}^2) (u +  \bar m_{2}^2 - m_{1}^2) + 2\, (t - m_{2}^2
-  \bar m_{2}^2) \,u\right],
\nonumber
\end{eqnarray}
where the coefficients $C_{...}$ depend on the channel, in
particular on isospin ($I$) and strangeness $(S)$. We refrain from
detailing the various coefficients since they can easily be
calculated and mostly have been given previously
\cite{GomezNicola:2001as}. The sums in (\ref{Tstu}) extend over
the members of the vector meson nonet, where $m_x$ denotes their
respective masses. The sum of the Mandelstam variables $ s+ t+ u =
m_1^2 + m_2^2 + \bar m_1^2 + \bar m_2^2 $ is the sum of the
squared meson masses, where we use a bar for the final state
masses. Partial-wave amplitudes are introduced by an average
\begin{eqnarray}
T^J(s) = \int_{-1}^{+1} \frac{d \cos \theta}{2} \,
\left(\frac{s}{\bar p_{\rm cm}\,p_{\rm cm}} \right)^J\,T(s,t,u)\,
P_J(\cos \theta) \,, \label{def-TJ}
\end{eqnarray}
over the center-of-mass scattering angle $\theta$. The relative
momenta of the initial and final state  are denoted by $p_{\rm
cm}$ and $\bar p_{\rm cm }$. The total angular momentum is $J$.

The coupled-channel partial-wave amplitudes $T^J_{ab}(s)$ are determined as
solutions of the non-linear integral equation
\begin{eqnarray}
T^J_{ab}(s)=U^J_{ab}(s)
+\sum_{c,d}\int^{\infty}_{\mu_{thr}^2}\frac{d\bar{s}}{\pi}\frac{s-\mu_M^2}{\bar{s}-\mu_M^2}\,
\frac{T^J_{ac}(\bar s)\,\rho^J_{cd}(\bar s)\,T^{J*}_{db}(\bar s)}{\bar{s}-s-i\epsilon}\,,
\label{def-non-linear}
\end{eqnarray}
with the phase-space matrix
\begin{eqnarray}\label{rho}
\rho^J_{ab} (s)=\frac{1}{8 \pi } \left(\frac{p_{cm }}{\sqrt{s}}\right)^{2\,J+1}\,\delta_{ab}\,.
\label{def-phase-space}
\end{eqnarray}
The matching scale $\mu_M^2$ is identified with the  smallest two-body threshold accessible in a
sector with isospin and  strangeness $(I,S)$.

Given a generalized potential $U^J_{ab}(s)$ the non-linear
integral equation implies a partial-wave scattering amplitude
$T^J_{ab}(s)$ that complies with the coupled-channel unitarity
request and the microcausality condition. On the other hand the
crossing symmetry constraint is not automatically satisfied
\cite{Gasparyan:2011yw}. If crossing symmetry would be implemented
exactly, a crossed reaction can be computed from the direct
reaction amplitude. Necessarily, the crossed reaction would be
determined by the direct reaction amplitude evaluated at energies
below the $s$-channel threshold. Now, imagine that in a first step
we have some means to approximate the generalized potential
systematically and very accurately at energies above the
$s$-channel threshold only. We may use this potential and
determine via (\ref{def-non-linear}) the scattering amplitude in
the physical region. In this case we are not able to compute the
crossed reaction, since our approximation for the generalized
potential is valid above threshold only, by assumption. The
crossed reaction may be computed by setting up the analogous
non-linear integral equation (\ref{def-non-linear}) for the
crossed reaction. It involves potentials distinct from the
potentials of the direct reaction. But where is the crossing
symmetry constraint? Our argument illustrates that crossing
symmetry is a constraint that affects dominantly amplitudes at
subthreshold energies. In a second step we now imagine that we
have some means to approximate the generalized potential
systematically at energies $\sqrt{s}\geq \Lambda_0  $, i.e. we
include a small subthreshold region. In this case crossing
symmetry does provide a constraint. The crossed reaction amplitude
can be computed from the direct amplitudes, however, only  in  a
specific subthreshold region. The coincidence of the crossing
transformed amplitudes and the amplitudes from the crossed
reaction in that specific subthreshold region defines the desired
constraint. Note that a non-empty coincidence region requires
$\Lambda_0$ to be sufficiently distinct from the $s$-channel
unitarity branch point. In our scheme a measure for the amount of
crossing-symmetry violation is the strength of non-perturbative
contributions to the subthreshold scattering amplitudes.

The non-linear integral equation~(\ref{def-non-linear}) does not necessarily have a
solution or it can have several. A unique solution is singled out
by the condition that in a small vicinity of the matching point
$\mu_M^2$, the results of chiral perturbation theory are
recovered. This implies a significant suppression of possible
crossing-symmetry violating contributions. From the form of (\ref{def-non-linear}) it follows that
the existence of a solution requires the generalized potential to
be bounded asymptotically, modulo some possibly logarithm terms.
Thus, an evaluation of $U^J_{ab}(s)$ in $\chi$PT is futile. Any
finite order truncation leads to unbounded potentials,
characterized by an asymptotic growth in some power of $s$.
However, as was pointed out in
\cite{Gasparyan:2010xz,Danilkin:2010xd}, the generalized potential
$U_{ab}^J(s)$ can be reconstructed unambiguously in terms of its
derivatives at a point $\mu_{ab,E}^2$ that lies within its
analyticity domain and where the results of $\chi$PT are reliable.
A solution of (\ref{def-non-linear}) requires
the knowledge of the generalized potential, $U^J_{ab}(s)$, for
real energies larger than the maximum of the initial and final
thresholds only. Following \cite{Gasparyan:2010xz} we identify
$\mu_E$ with the mean of initial and final thresholds. A Taylor
expansion of $U(s)$ around $\mu_E^2$ has a small convergence
radius, that is determined by the distance to the closest
left-hand cut. In our case the closest left-hand cuts reflect
the two-body t- and u-channel unitarity condition. In order to
extend the convergence up to some cutoff scale $\Lambda_s$, we
apply the conformal map, that was constructed in
\cite{Gasparyan:2010xz}. Explicitly, it is given by
\begin{eqnarray}
\xi(s)=\frac{a\,(\Lambda^2_s-s)^2-1}{(a-2\,b)(\Lambda^2_s-s)^2+1}\,,\quad a = \frac{1}{(\Lambda^2_s-\mu_E^2)^2},
 \quad b = \frac{1}{(\Lambda^2_s-\Lambda^2_0)^2}\,,
 \label{def-conformal}
\end{eqnarray}
where the parameter $\Lambda_0$ is identified unambiguously such that the
mapping domain of the conformal map touches the closest left-hand branch point. Within this domain the generalized
potential can be reconstructed in terms of its derivative at the expansion point $\mu_E^2$. It holds
\begin{eqnarray}\label{U expansion}
U(s )=\sum_{k=0}^{\infty} \,c_k \,\xi^k(s) \qquad {\rm for} \quad s < \Lambda_s^2 \,,
\label{def-U-xipanded}
\end{eqnarray}
where the coefficient $c_k$ is determined by the first $k$ derivatives of $U(s)$ at $\mu_E^2$.
In our analysis the coefficient $c_k$  are all determined from (\ref{Tstu}) by the masses of the pseudo-scalar and
vector mesons and the two parameters $f$ and $h_P$. It is emphasized that given our construction any contribution, be it
a tree-level or loop term, contributes to the
generalized potential exclusively via its derivatives at the expansion point. All left-hand cut structures
are 'integrated' out systematically. In this work we construct the generalized potential $U^J_{ab}(s)$ in terms of
its first few derivatives at  the expansion point $\mu_E^2$, where we use the derivatives implied by the tree-level
amplitude (\ref{Tstu}). Via (\ref{def-U-xipanded}) we obtain an approximate generalized potential for energies
$\Lambda_0 < \sqrt{s} < \Lambda_s$. For energies larger than the cutoff scale $\Lambda_s$ the generalized potentials
are set to a constant \cite{Gasparyan:2010xz}. By virtue of the specific form of the conformal map this is a smooth procedure.

It remains to find the desired solution of (\ref{def-non-linear}).
This is readily achieved in application of the $N/D$ technique
\cite{Chew:1960iv}. The partial-wave scattering amplitude is
decomposed as
\begin{eqnarray}
T_{ab}(s)=\sum_{c}\,D^{-1}_{ac}(s)\,N_{cb}(s)\,.
\label{def-NoverD}
\end{eqnarray}
Again, contributions of right- and left-hand cuts are separated. $D_{ab}(s)$ contains only the right-hand
s-channel unitarity cuts. With the ansatz
\begin{eqnarray}
&&D_{ab}(s)=\delta_{ab} -\frac{s-\mu^2_M}{s-M^2_{CDD}}\,R^{(D)}_{ab}
 -\sum_{c}\int_{\mu_{thr}^2}^{\infty}\frac{d\bar{s}}{\pi}\frac{s-\mu_M^2}{\bar{s}-\mu_M^2}
\frac{N_{ac}(\bar{s})\rho_{cb}(\bar{s})}{\bar{s}-s}\,,
\label{def-D}
\end{eqnarray}
the coupled-channel unitarity is ensured~\cite{Gasparyan:2010xz}.
In (\ref{def-D}) we allow for one CDD pole structure
\cite{Castillejo:1955ed} with its pole mass parameter $M_{CDD}$
and coupling matrix $R^{(D)}$. Such a term is requested by the
presence of the s-channel vector meson exchange process in some
$J=1$ amplitudes \cite{Oller:1998zr,Szczepaniak:2010re}. The
matrix $N_{ab}(s)$ contains left-hand cuts only. If the ansatz
(\ref{def-NoverD}) is to represent a solution of the non-linear
equation (\ref{def-non-linear}) the matrix function $N_{ab}(s)$
has to satisfy the linear integral equation
\begin{eqnarray}
&& N_{ab}(s)=U^{\rm eff}_{ab}(s)
-\frac{s-\mu^2_M}{s-M^2_{CDD}}\,\Big[R^{(B)}_{ab}+\sum_c R^{(D)}_{ac}\,U^{\rm eff}_{cb} (s)\Big]
\nonumber\\
&&\qquad  \;\,+\,\sum_{c,d}\int_{\mu_{\rm thr}^2}^\infty \frac{d\bar s}{\pi}\,\frac{s-\mu^2_M}{\bar s-\mu^2_M}\,
\frac{N_{ac}(\bar s)\,\rho_{cd}(\bar s)\,[U^{\rm eff}_{db}(\bar s)-U^{\rm eff}_{db}(s)]}{\bar s-s}\,,
\label{CDD-ansatz}
\end{eqnarray}
with
\begin{eqnarray}
&& U^{\rm eff}_{ab}(s) = U_{ab}(s) + \frac{g_a\,m^2\,g_b}{s-m^2}\,\frac{s-\mu^2_M}{m^2-\mu^2_M}\,,
\nonumber\\
&& R^{(D)}_{ab}=\frac{m^2-M_{CDD}}{m^2-\mu^2_M}\left(\delta_{ab}-\sum_c\,\int_{\mu_{\rm thr}^2}^\infty \frac{d\bar s}{\pi}\,\frac{m^2-\mu^2_M}{\bar s-\mu^2_M}\,\frac{N_{ac}(\bar s)\,\rho_{cb}(\bar s)}{\bar s-m^2}\right)\,,
\nonumber\\
&& R^{(B)}_{ab}=-\frac{\mu^2_M-M_{CDD}^2}{(\mu^2_M-m^2)^2}\,g_a\,m^2\,g_b
\nonumber\\
&&\qquad \,-\,\sum_{c,d}\,\int_{\mu_{thr}^2}^\infty \frac{d\bar s}{\pi}\,\frac{\bar s-M_{CDD}}{\bar s-\mu^2_M}\,
\frac{N_{ac}(\bar s)\,\rho_{cd}(\bar s)}{(\bar s-m^2)^2}\,g_d\,m^2\,g_b\,
\nonumber\\
&&\qquad \,+\,(m^2-M_{CDD}^2)\,\sum_{c,d}\,\int_{\mu_{ \rm thr}^2}^\infty \frac{d\bar s}{\pi}\,
\frac{N_{ac}(\bar s)\,\rho_{cd}(\bar s)\,U^{\rm eff}_{db}(\bar s)}{(\bar s-\mu^2_M)\,(\bar s-m^2)}\,.
\label{identify-R}
\end{eqnarray}
The merit of the ansatz (\ref{CDD-ansatz}, \ref{identify-R}) lies
in its specification of the CDD pole parameters, $R^{(D)}$ and
$R^{(B)}$,  in terms of the parameters, $g_a, m$, characterizing a
possible pole term in the generalized potential. They have to be
dialed such that the effective potential $U^{\rm eff}_{ab}(s)$ is
regular at $s= m^2$. The CDD pole mass parameter, $M_{CDD}$, is
irrelevant. By construction, the scattering amplitude, which
results from (\ref{def-NoverD}-\ref{identify-R}), does not depend
on the choice of $M_{CDD}$.

The system (\ref{CDD-ansatz}, \ref{identify-R}) can be solved numerically by matrix inversion techniques.
Once we obtain a solution of $N_{ab}(s)$, we can compute $D_{ab}(s)$ via (\ref{def-D}), and  a well defined
result for the partial-wave scattering amplitude is obtained with (\ref{def-NoverD}).

\section{Numerical results}

Our results rely on the choice of various parameters. We use $f =
90$ MeV in the following, where we checked that small variations
around that value lead to very similar results. For the cutoff
parameter $\Lambda_s$ introduced in the conformal variable
(\ref{def-conformal}) we use the $\Lambda_s = 1.4$ GeV unless
stated otherwise. The small effects caused by variations around
that central value will be discussed. The value of $\Lambda_s$
sets the scale from where on s-channel physics is integrated out.
In our approach the generalized potential is characterized further
by the number of terms consider in the conformal expansion
(\ref{def-U-xipanded}). Typically a few terms only suffice to
arrive at a good approximation of the generalized potential. A
large number of terms would be justified if the left-hand cut
structures are modeled very accurately, or a large number of
counter terms are considered. If the empirical data set would be
very accurate and more complete one could determine the various
coefficients in (\ref{def-U-xipanded}) directly from the data set
in a model independent manner. In our approach the left-hand cut
structures are defined by the t- or u-channel exchange of
Goldstone boson pairs. The latter imply specific correlations of
the expansion coefficients.

\begin{figure}[t]
\includegraphics[height=6.5cm]{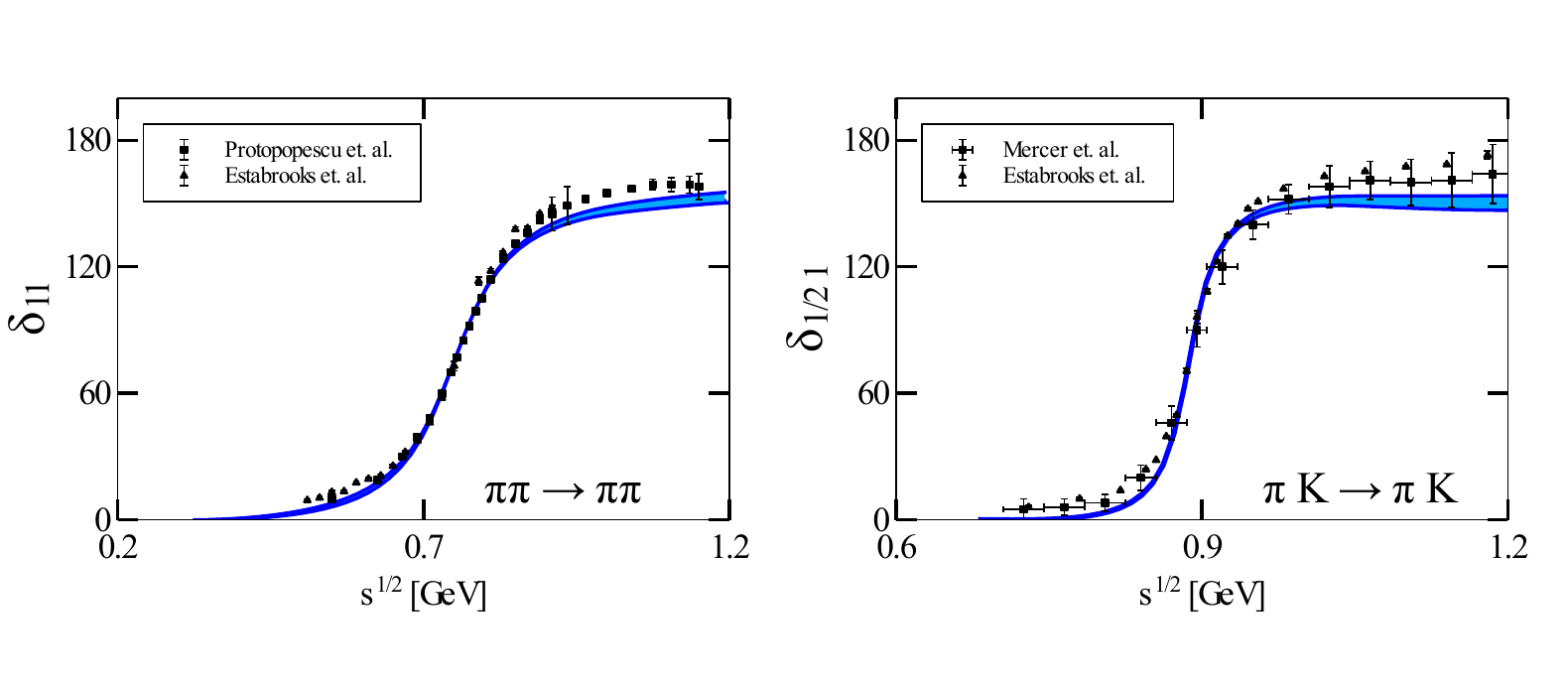}
\caption{p-Wave scattering phases $\delta$ for $\pi \pi \to \pi
\pi $ with $I=1$ and $\pi K \to \pi K$ with $I=\frac{3}{2}$. The
data are taken form
\cite{Protopopescu:1973sh,Estabrooks:1974vu,Mercer:1971kn,Estabrooks:1977xe}.}
%\label{fig:1}
\end{figure}

We begin with a presentation of partial-waves with $J=1$, which probe the masses and decay properties of the light
vector mesons. At tree-level the vector meson masses are
\begin{eqnarray}
&& m_{\rho}^2= m_{\omega}^2=m_{1^{-}}^2+b_D \,m_{\pi}^2\,,\qquad  m_{K^*}^2=m_{1^{-}}^2+b_D \,m_{K}^2\,,
\nonumber\\
&& m_{\phi}^2=m_{1^{-}}^2+b_D \,(2\,m_{K}^2-m_{\pi}^2)\,,
\label{def-vector-masses}
\end{eqnarray}

which leads to the estimate $m_{1^-}\simeq 0.76$ GeV and $b_D
\simeq 0.95$. The  latter values reproduce the empirical vector
meson masses with an uncertainty of less than 10 MeV. Within our
unitarization scheme we identify the expressions of
(\ref{def-vector-masses}) with the mass parameter $m$ in
(\ref{identify-R}). The physical $\omega$ and $\phi$ meson masses
can be extracted from the $(I^G,S)=(0^-,0)$ sector, which involves
the single channel $\bar K K$ only. The solution of
(\ref{def-NoverD}-\ref{identify-R}) recovers the tree-level mass
of the $\omega$ meson identically. Since its mass is below the
$\bar K K$ threshold, its decay width is zero here. In contrast
the $\phi$ meson receives a small width depending on the actual
value of $h_P$. Its mass is almost indistinguishable from its
tree-level expression (\ref{def-vector-masses}). We determine $h_P
\simeq 0.26$ as to obtain an accurate description of the $(I,S) =
(1,0)$ and $(I,S)=(1/2,1)$ sectors, which are dominated by poles
caused by the $\rho$ and $K^*$ mesons. We find that the mass
parameters $m$ in (\ref{identify-R}) needed to recover the
empirical $\pi \pi$ and $\pi K$ phase shifts are the empirical
masses of the $\rho$ and $K^*$ mesons. We conclude that the
tree-level expressions for the vector meson masses survive the
unitarization. This justifies our procedure to use physical vector
meson masses in the driving expressions (\ref{Tstu}). In Fig.1 we
confront our results with empirical p-wave $\pi \pi$ and $\pi K$
phase shifts. A good description up to about 1 GeV is achieved.
The $\rho$ and $K^*$ mesons are clearly visible in the data and
our theoretical curves. As a consistency check we computed with
$\Gamma_\phi \simeq 3.47$ MeV the width of the $\phi$ meson. Our
value is consistent with the empirical range of the partial width
$\Gamma_\phi(K \bar K) = 3.54\pm0.04$ MeV. It is amusing to
observe that a universal value of $h_P$ is able to recover the
decay properties of the $\rho$, $K^*$ and $\phi$ mesons
simultaneously.

\begin{figure}[t]
\begin{center}
\vskip-0.1cm \hskip-0.5cm
\parbox{14.0cm}{\includegraphics[clip=true,width=14.0cm]{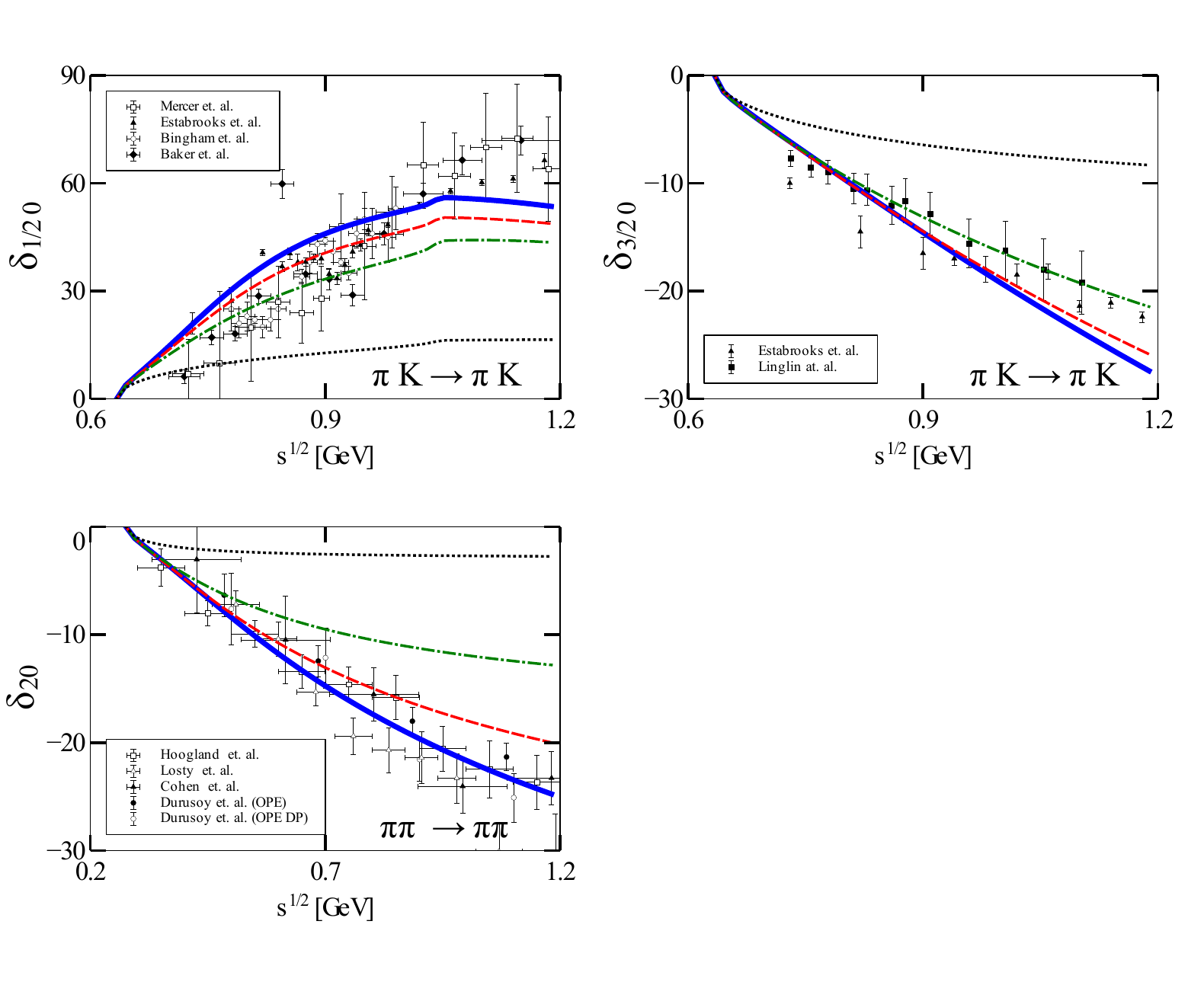}}
\parbox{13.5cm}{\vspace{-6.0cm}
\hspace*{7.3cm} \parbox{6cm}{\caption{s-Wave scattering phases for
$\pi K \to \pi K$ with $I=\frac{1}{2},\frac{3}{2}$ (upper panel)
and $\pi \pi \to \pi \pi$ in $I=2$ (lower panel). The dotted,
dash-dotted, dashed and solid lines correspond to a truncation at
order $0,\ 1,\ 2,\ 3$ in the expansion (\ref{def-U-xipanded})
respectively. The data are taken from
\cite{Mercer:1971kn,Estabrooks:1977xe,Bingham:1972vy,Baker:1974kr,Linglin:1973ci,Hoogland:1977kt,Losty:1973et,Cohen:1980cq,Durusoy:1973aj}.}}}
\end{center}
%\label{fig:2}
\end{figure}

Our results are stable against variations of $\Lambda_s$ and
changing the number of terms kept in (\ref{def-U-xipanded}). A
variation of $\Lambda_s$ from 1.1 GeV to 1.6 GeV would almost not
be visible in the figure and therefore we refrain from showing it.
In contrast decreasing the number of terms in
(\ref{def-U-xipanded}) from four to one defines the bands shown in
Fig.1.

We turn to a discussion of the s-waves, where we again focus on
phases for which there is empirical information available. Our
results for the $\pi K$ phases are shown against empirical data in
Fig.2. The figure includes also our result for the isospin two
s-wave $\pi \pi$ scattering phase. While a variation of
$\Lambda_s$ from 1.1 GeV to 1.6 GeV would barely be visible in the
figure, the results are sensitive to the numbers of terms
considered in (\ref{def-U-xipanded}). The figure shows the results
of keeping 1 to 4 terms in (\ref{def-U-xipanded}). In all three
cases there is a saturation observed, with already the 4-term
approximation recovering the empirical phases within the
uncertainties of the different experimental results. We conclude
from this result that indeed the vector meson exchanges
contributions provide a quite accurate approximation for the
leading coefficients in the conformal expansion
(\ref{def-U-xipanded}).

\begin{figure}[t]
\begin{center}
\vskip-0.1cm \hskip-0.5cm
\parbox{14.0cm}{\includegraphics[clip=true,width=14.0cm]{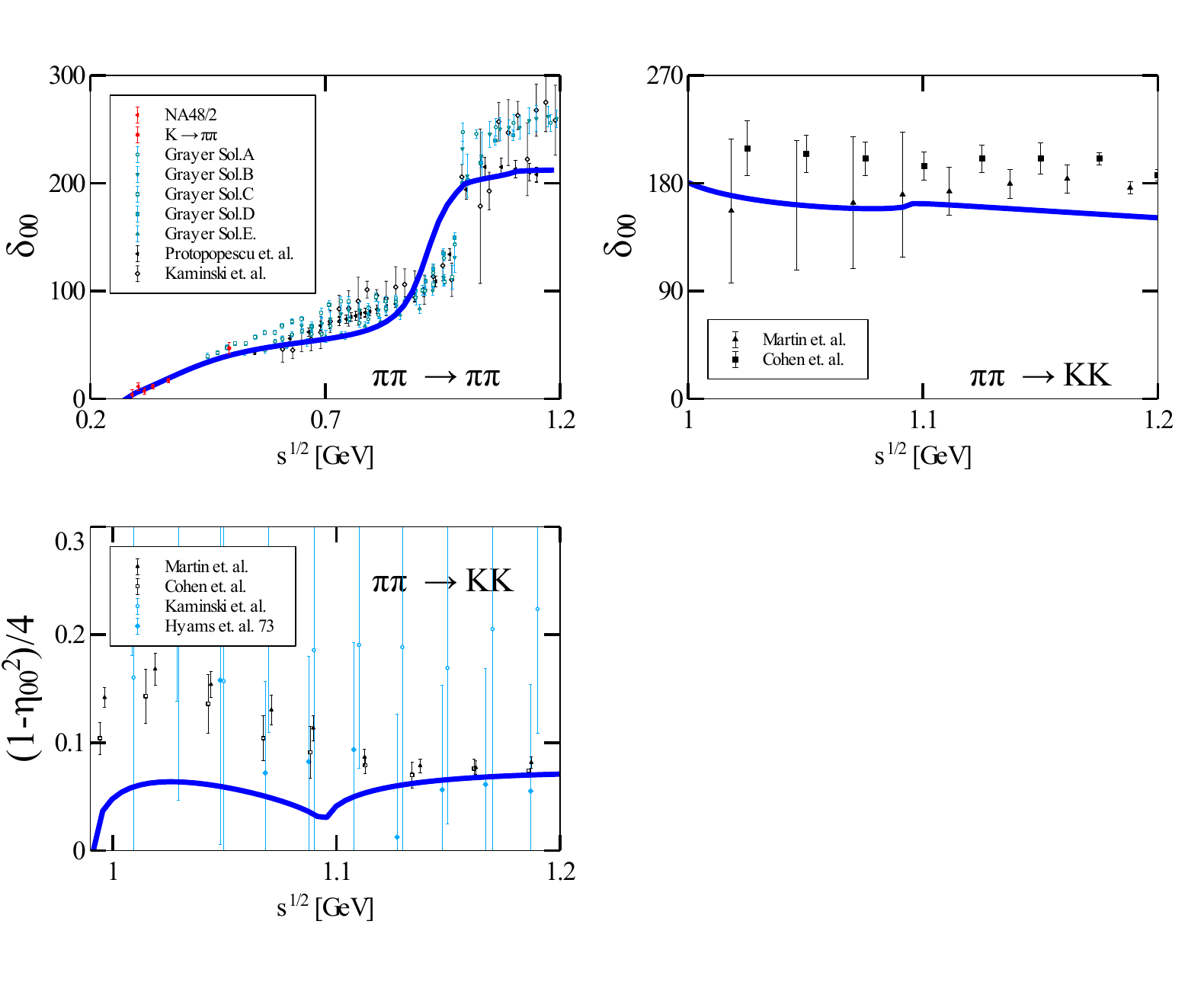}}
\parbox{13.5cm}{\vspace{-6.0cm}
\hspace*{7.3cm} \parbox{6cm}{\caption{s-Wave scattering phases for
$\pi \pi \to \pi \pi $ and $\pi \pi \to \bar K K  $  with  $I=0 $
(upper panel) and the inelasticity $\pi \pi \to \bar K K  $ (lower
panel). The solid lines correspond to truncation the expansion
(\ref{def-U-xipanded}) with 4 terms. The data are taken from
\cite{Grayer:1974cr,Protopopescu:1973sh,Kaminski:1996da,Batley:2007zz,Cirigliano:2009rr,Martin:1979gm,Cohen:1980cq,Kaminski2001,Hyams:1973zf}.}}
}
\end{center}
%\label{fig:3}
\end{figure}

We close the presentation with a discussion of the isoscalar
sector. Three different channels, $\pi\, \pi$, $\bar K\, K$ and
$\eta \,\eta$ contribute in our analysis. In Fig.3 the empirical
data from $\pi \pi$ scattering are collected and confronted
against our results. The solid lines follow in the 4-term
approximation. We recover the $\pi \,\pi \to \pi\, \pi$ and $\pi
\,\pi \to \bar K \, K$ phases qualitatively, with some
quantitative discrepancies around the $f_0(980)$ resonance
structure. The inelasticity in the $\pi \pi \to \bar K K$ reaction
appears to be underestimated.

Our results are in reasonable agreement with previous results from
the Roy-Steiner equations \cite{Ananthanarayan:2000ht,Colangelo:2001df,Buettiker:2003pp}.
With our analysis we further explored the role of explicit vector-meson
degrees of freedom. Our leading order computation has basically only two free parameters
$f=90$ MeV and $h_P=0.26$. To this extend we find the agreement
with the previous results based on Roy-Steiner equations \cite{Ananthanarayan:2000ht,Colangelo:2001df,Buettiker:2003pp}
the very satisfactory.

\section*{Acknowledgements}
We acknowledge useful discussions with A.M. Gasparyan.

\bibliography{Igorbib}{}
\bibliographystyle{elsart-num}

\end{document}